\documentclass{aa}
\usepackage{graphicx}
\usepackage{color}

\begin{document}      
   
   \title{Quasi-simultaneous multi-frequency observations of inverted-spectrum GPS candidate sources \thanks{Tables 5 and 6 are only available in electronic form at the CDS via anonymous ftp to cdsarc.u-strasbg.fr (130.79.128.5) or via http://cdsweb.u-strasbg.fr/cgi-bin/qcat?J/A+A/.}}

   \author{B.~Vollmer\inst{1}, T.P.~Krichbaum\inst{2}, E.~Angelakis\inst{2}, Y.Y.~Kovalev\inst{2,}\inst{3}}

   \offprints{B.~Vollmer: bvollmer@astro.u-strasbg.fr}

   \institute{CDS, Observatoire astronomique de Strasbourg, 11, rue de l'universit\'e,
              67000 Strasbourg, France \and
              Max-Planck-Insitut f\"{u}r Radioastronomie, Auf dem H\"{u}gel 69, 53121 Bonn, Germany \and
              Astro Space Center of Lebedev Physical Institute, Profsoyuznaya 84/32, 117997 Moscow, Russia}

   \date{Received / Accepted}

   \authorrunning{Vollmer et al.}
   \titlerunning{Quasi-simultaneous observations of inverted-spectrum radio sources}

\abstract
{
Gigahertz-Peaked Spectrum (GPS) sources are probably the precursors of local radio galaxies.
Existing GPS source samples are small ($<200$).
}
{
It is necessary to extend the availabe sample of the Gigahertz-Peaked
Spectrum (GPS) and High Frequency Peaker (HFP) sources in order
to study their nature with greater details and higher statistical significance.
}
{
A sample of 214 radio sources, which were extracted from the SPECFIND
catalog and show an inverted radio spectrum, were observed
quasi-simultaneously at 4.85, 10.45, and 32~GHz with the 100-m
Effelsberg radio telescope. Using the VLBA calibrator survey (VCS) we
have investigated the parsec-scale morphology of the sources. 
}
{
About 45\% of the sources in our sample are classified as GPS or HFP
candidates. We add 65 new GPS/HFP candidates to existing samples.
We confirm the expected tendency that HFP are more compact on
milliarcsecond scale than the 'classical' GPS sources, which peak at lower frequencies.
}
{
The data mining of the SPECFIND database
represents a promising tool for the discovery of new GPS/HFP sources.
}

\maketitle

\color{black}

\section{Introduction \label{sec:introduction}}

GHz-Peaked Spectrum (GPS) sources are powerful ($\log
P_{1.4~{\rm GHz}} > 25$~W\,Hz$^{-1}$) and compact ($< 1$~kpc)
extragalactic radio sources, which show a convex radio spectrum
peaking between 500~MHz and 10~GHz in the observer's frame (for a
review see O'Dea~1998).
The physical mechanism responsible for the turnover of the spectrum is
still unclear with two competing models proposed:
the synchrotron self-absorption caused by dense
plasma within the source or the free-free
absorption caused by a screen external to the source.

High Frquency Peakers (HFP) are radio sources defined via their convex
spectrum peaking at frequencies above 5~GHz (Dallacasa et al. 2000).
From the anti-correlation found between the turnover frequency
and size (O'Dea \& Baum 1997) HFPs are expected to be smaller and therefore 
younger radio sources than GPS sources.

GPS sources are associated with either quasars or galaxies. Despite
the similar shape of their radio spectrum, these two classes of GPS
sources are often considered to be different. 
Torniainen et al. (2005), who studied the long term variability of 35
inverted-spectrum sources, concluded that genuine quasar-type GPS
sources are rare. They found a large number of highly variable blazar sources
that can have a convex spectrum peaking at high frequencies (up to
$\sim$100~GHz) during flaring events occurring in the radio jets.

The nature of GPS 
sources is still under debate. Two possible scenarios have been put
forth: (i) the `frustration' scenario, according to which the small
size and the inverted spectrum are caused by a dense environment that
prohibits the source
from growing larger (e.g., Gopal-Krishna \& Wiita 1991), (ii) the
`youth' scenario, suggesting that the GPS sources represent the young
precursors of compact steep spectrum (CSS) sources and extended radio
sources (e.g., Mutel \& Phillips 1988, Fanti et al. 1990, 1995).

There is now a wide consensus
that, at least the symmetric GPS sources, represent the early evolutionary
stage of the extended radio source population. At this stage the radio
emitting region grows and expands within the interstellar medium
before plunging into the intergalactic medium to form an FR~II radio
source (Fanti et al.~1995, Readhead et al.~1996, Begelman~1996,
Snellen et al.~2000a).

The detection of kpc-scale emission associated with a few GPS
sources seems to be inconsistent with a recent origin of the radio activity.
Such an extended emission is interpreted as a sign of a past nuclear activity. In this case
the GPS source, i.e. the galactic nucleus, is at the beginning of a new cycle of activity.
Since extended radio emission around GPS sources is
a rare phenomenon, Stanghellini et al. (2005) conclude that
the time scale between subsequent phases of activity is in general longer
than the radiative lifetime of the radio emission from the earlier activity
($\sim 10^{8}$~yr).

The currently existing GPS sample is limited ($<200$ objects, see, e.g.,
the recent compilation by Labiano et al.~2007). 
In order to conduct statistical studies of GPS objects and test whether the number
of GPS sources at intermediate redshifts is in accordance with that of
local FR~II radio galaxies, it is necessary that the sample of
available GPS sources is extended.
In particular, this could help to test whether GPS sources are the precursors
of the local FR~II radio galaxies.

In this paper, we present results of a search for new GPS/HFP candidates.
Since the identification of a radio source as a genuine young GPS source
depends on the detailed knowledge of its spectrum, flux density variability, VLBI structure
and source identification, we will use in the following the words GPS or HFP source as a synonym 
for GPS candidate or HFP candidate.
In order to measure radio-spectra over a range of high frequencies, which are
not affected by non-contemporanous measurements, we observed a sample of 214 objects north of declination
$-25^\circ$ that show an inverted radio spectrum,
with quasi-simultaneous flux density measurements at 4.85~GHz,
10.45~GHz, and 32~GHz using the Effelsberg 100-m telescope of the
Max-Planck-Institut f\"ur Radioastronomie (MPIfR).
In Sect.~\ref{sec:sample}
we review the sample selection procedure implemented on the basis of the
SPECFIND database, observations and primary data reduction are described
in Sect.~\ref{sec:observations}.
We present results of
observations in Sect.~\ref{sec:results},
sources extention and variablility characteristics in Sect.~\ref{sec:ext_var},
cross-identification and spectral fitting in Sect.~\ref{sec:fitting},
milliarsecond-scale compactness and morphology in Sect.~\ref{sec:vlbi},
comparison with other available GPS/HFP samples in Sect.~\ref{sec:discussion}.
We summarize our results in Sect.~\ref{sec:summary}.

\section{Sample selection\label{sec:sample}}

\begin{figure}[b]
        \resizebox{\hsize}{!}{\includegraphics[trim=1cm 0cm 0.2cm 0cm]{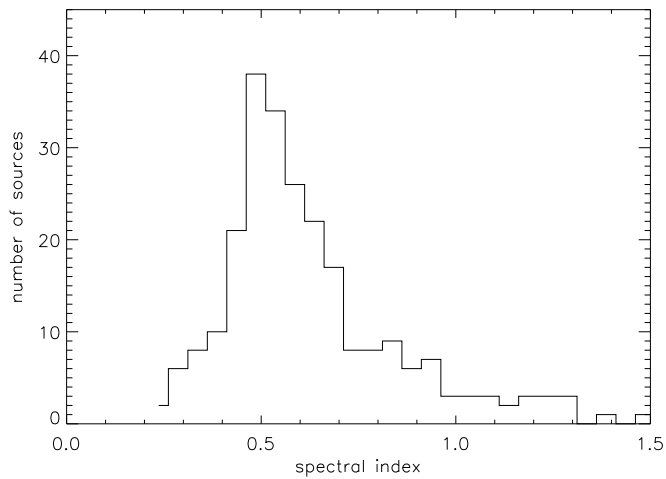}}
        \caption{SPECFIND spectral index
        distribution of the sources in our sample.  The power law
        spectrum is defined by $S_{\nu}\propto \nu^{\alpha}$.  All the
        objects have several spectral indices depending on the
        weighting of the frequency points. Objects with $\alpha < 0.5$
        have at least one associated spectral index $\geq 0.5$.  }
        \label{fig:si_dist}
\end{figure}

SPECFIND (Vollmer et al. 2005a) is a hierarchical algorithm. It classifies a source $j$ as
parent, sibling, or child with respect to a given source $i$ using a
procedure which has different stages:\\ stage 1: cross-identification
based on proximity criteria taking into account the resolution and
source size,\\ stage 2: cross-identification based on flux densities
at the same frequency,\\
stage 3: cross-identification based on flux density at different
frequencies, as expected from the radio spectral index, adopting the convention 
$S_{\nu}\propto \nu^{\alpha}$.  

At the end of this procedure a source $i$ and its siblings are
considered to be the same source.  If source $j$ is identified as a
parent, source $i$ might be a resolved sub-source of source $j$.  If
source $i$ has children, it might be extended, and the children
represent its resolved sub-sources.  At the end of the
cross-identification the self-consistency of the hierarchy and the
uniqueness is tested. A radio source cannot be assigned to more than
one physical object.  The position of each object is taken from the
radio source observed with the highest resolution (for the positional
accuracy of the included radio surveys see Vollmer et al. 2005b).

The result of the cross-identification depends strongly on the
detailed spectrum fitting algorithm (Vollmer et al.~2005a), the
initial set of frequency points and their weights. During this
procedure non-fitted frequency points are rejected and removed from
the spectrum, since they may belong to a different object.  SPECFIND
determines the slope and the abscissa for a given source $i$ with all
other cross-identified sources. It is required that the final power
law fits source $i$ within the errors. Therefore, the result of the
fitting algorithm can be different depending on the choice of source
$i$.  A given object with an associated radio spectrum can thus
possess several slopes and abscissa. 

From an extended version of the SPECFIND catalog\footnote{available
at CDS VizieR (Ochsenbein et al. 2000)}, which contains the
comparison samples described in table 5 of Vollmer et al. (2005a),
we extracted objects with a maximum slope (spectral index, $\alpha$)
greater or equal 0.5 and declination $\delta > -25^\circ$. 
The frequency range in which the spectral index is determined 
depends on the actual frequency coverage and sensitivity of the 22 individual radio catalogues
used by SPECFIND. To avoid
confusion problems for the flux density measurements performed at Effelsberg, we then removed
all sources which are extended in the NVSS (larger than the $45"$ beam size;
Condon et al. 1998) and
some galactic sources, which coincide with known H{\sc ii} regions.

\begin{figure}[b]
        \resizebox{\hsize}{!}{\includegraphics[trim=1cm 0cm 0.2cm 0cm]{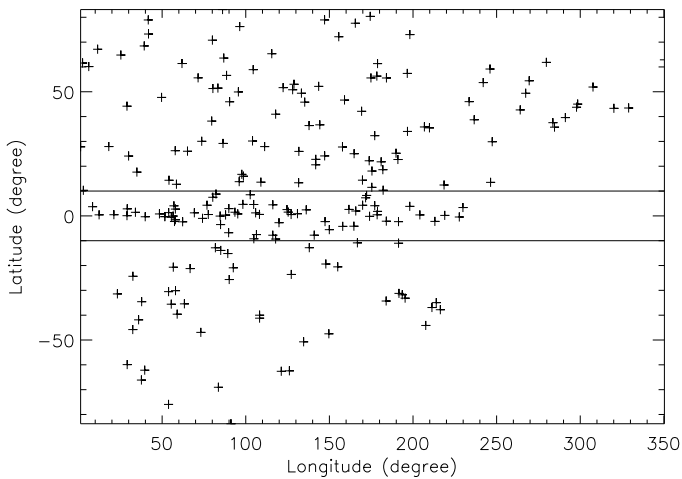}}
        \caption{Spatial distribution of our
        sample of 214 sources with an inverted radio spectrum in
        Galactic coordinates . The horizontal lines mark galactic
        latitudes of $\pm 10^{\circ}$.}
        \label{fig:gpsplot}
\end{figure}
This resulted in a list of 214 objects. The majority of the sources
has a spectral index of around $+0.5$.  The distribution of spectral indices
of these sources is shown in Fig.~\ref{fig:si_dist}, and their
distribution on the sky is shown in Fig.~\ref{fig:gpsplot}.  About one
third of them (69) are located within $\pm 10^{\circ}$ from the
Galactic plane. Some of them were later identified as infrared sources
or planetary nebula.  Since our selection criterion is only based on
the shape of the radio spectrum, we did not remove these sources. The
number of targets is smaller in the southern hemisphere than in the
northern hemisphere because of the lack of deep surveys in the
southern hemisphere (Vollmer et al. 2005a).

\section{Observations and primary data reduction\label{sec:observations}}

The flux density measurements of the $214$ target sources were made
with the 100-m MPIfR radio telescope in Effelsberg (Germany). The
observations took place in 6 observing sessions which are summarized
in Table~\ref{tab:sessions}.

In each session the target
sources and the calibrator sources (see Table~\ref{tab:calib}) were
observed with the heterodyne receivers mounted on the secondary focus
of the telescope.  The measurements were made at 6\,cm (4.85\,GHz),
2.8\,cm (10.45\,GHz) and 9\,mm (32\,GHz). The receiver characteristics
are summarized in Table~\ref{tab:receiver}. Column 1 gives the
observing wavelength, column 2 the central frequency, column 3 the
observing bandwidth, column 4 the typical system temperature (at
zenith), column 5 the typical peak value of the elevation-dependent
antenna gain, column 6 the size of the observing beam
(full width at half maximum, FWHM), column
7 the polarization of the receiver feeds, column 8 the number of
receiver horns used in the data reduction\footnote{the 6-cm receiver
has 2 feeds, the 2.8-cm one has 4 and the 9\,mm receiver has 6 feeds},
and column 9 the receiver type. The signal from the second receiver
horn allowed the off-source signal containing the (time variable)
atmospheric signal to be subtracted from the signal on the source
(beam switch). This improved the data quality and led to almost flat
baselines in each scan.

\begin{table}[t]
\caption{Effelsberg observing sessions \label{tab:sessions}}
\center
\begin{tabular}{rc}
\hline\hline
Date            &UT range (hr)\\
\hline
23, 24 Sep 2005 & 11--08 \\
 8, 9  Oct 2005 & 07--04 \\
17, 18 Jan 2006 & 18--02 \\
24, 25 Jan 2006 & 16--08 \\ 
30, 31 Jan 2006 & 15--08 \\
    24 Mar 2006 & 02--10 \\
\hline
\end{tabular}
\end{table}

For each source, the antenna temperature was measured using the method
of 'cross-scans'. Here, the telescope beam is repeatedly moved across
the source position, with an equal number of sub-scans in elevation
and in azimuth direction.  The number of sub-scans in each cross-scan
ranged between $4-12$, depending on the flux density of the source and
the stability of the telescope pointing. Each source was measured with
at least two cross-scans at each of the three observing bands. With a
typical duration of 25-30 seconds for each sub-scan and scan lengths
of $600''$, $220''$, and $110''$ at $\lambda=$ 6\,cm, 2.8\,cm, \&
9\,mm, respectively, each scan lasted between $3-5$\,min. This allowed
us to obtain for each source `quasi-simultaneous' flux density
measurements at the 3 observing bands within a $15-20$\,min time
interval.

After the observations, the data were reduced in the standard manner
following the procedures described in Kraus et al.~(2003). 
To derive the flux densities of the observed point sources, Gaussian
profiles were fitted to each sub-scan in each driving direction. For
fainter sources, the sub-scans were averaged in each driving direction
before the Gaussian fitting was done. After correcting for small
residual pointing offsets, the amplitudes of the individual sub-scans
were combined by averaging them. For each scan and source this
resulted in the uncalibrated antenna temperature as determined by the
peak value, $S^\mathrm{peak}$, of the Gaussian fit. For each scan,
also the half-power beam width, $HPBW^\mathrm{obs}$, was measured. In
the next step we corrected the measurements for the
elevation-dependent antenna gain and for systematic time-dependent
effects, using standard gain-elevation curves and the known flux
densities of the primary and secondary calibrators, which where
observed regularly at $3-6$\,hr intervals throughout each of the
observing sessions. Opacity corrections were applied to the 9\,mm data.
The conversion from antenna temperature (measured
in K) to flux density (measured in Jy) was done using the flux density
scale of Ott et al.~(1994), refined by more recent measurements
(A.~Kraus, priv.~comm.).  Table~\ref{tab:calib} summarizes the flux
density scale used throughout this paper.

\begin{table}
\caption{The primary and the secondary calibrators along with and
their flux densities at the observing frequencies.
\label{tab:calib}}
\begin{tabular}{lcccl}
\hline\hline
Source  &$S_{\rm 6 cm}$ & $S_{\rm 2.8 cm}$ &$S_{\rm 9 mm}$&  Comment               \\
        &[Jy]           &     [Jy]         &    [Jy]      &                       \\
\hline
3C48    &5.48           &    2.60          &   0.80       &Primary Calibrator    \\
3C138   &3.79           &    2.16          &   0.92       &Secondary Calibrator    \\
3C147   &7.90           &    3.82          &   1.18       &Secondary Calibrator    \\
3C161   &6.62           &    3.06          &   0.83       &Secondary Calibrator    \\
3C286   &7.48           &    4.45          &   1.82       &Primary Calibrator    \\
3C295   &6.56           &    2.62          &   0.55       &Primary Calibrator    \\
NGC7027 &5.48           &    5.92          &   5.49       &Primary Calibrator    \\
\hline
\end{tabular}
\end{table}

\begin{table*}
\caption{The characteristics of the used receivers.}
\label{tab:receiver}
\begin{tabular}{ccccccccc}
\hline\hline
$\lambda$  &  Center $\nu$     & Bandwidth & $T_{sys}$ &  Gain    & FWHM &  Polarization & Horns &  Comment \\
~[cm]      &    [GHz]          &  [MHz]    &    [K]    &  [K/Jy]  &   ["]  &               &       &          \\
\hline
6          &     4.85          &   500     &   27      &   1.5    &   146  &  dual circ.   &   2   &  Software Beam switch \\
2.8        &    10.45          &   300     &   50      &   1.3    &    67  &  dual circ.   &   2   &  Software Beam switch \\
0.9        &    32.00          &  2000     &   77      &   0.5    &    26  &  LCP          &   2   &  Correlation Receiver \\
\hline 
\end{tabular}
\end{table*}

\section{Results \label{sec:results}}

The observed radio flux densities at 4.85, 10.45, and 32~GHz are
listed in Table~\ref{tab:table} along with the spectral indices
between 4.85-10.45~GHz and 10.45-32~GHz.  
The last column of the table lists the observing frequency in GHz, at
which the FWHM of the Gaussian fitted to the telescope response was
significantly larger than that to the point
source (see Table~\ref{tab:receiver}). This is indicative of extended
source structure or confusion.
The sources marked in this way should be regarded either as partially
resolved or as being confusing within the Effelsberg observing beam.
We detected all 214 sources at 4.85~GHz, 209 (98\,\%) sources at 10.45~GHz, and
181 (85\,\%) sources at 32~GHz.  For the non-detected sources we assign a dash
('$-$') in the corresponding column of Table~\ref{tab:table}.
Depending on the on-source integration time, receiver properties and
weather conditions the estimated detection limits are $3\sigma=1-5$\,mJy at
4.85\,GHz, $3-10$\,mJy at 10.45\,GHz and $10-30$\,mJy at 32\,GHz, with
the higher numbers of each range reflecting non-optimal weather
conditions.
The provided errors of the flux density measurements are $1\sigma$
errors and result from formal error propagation through the following
steps: Gaussian fitting, gain and time correction, K/Jy conversion.
The error of the absolute flux density scale (Baars
et al.~1977, Ott et al.~1994) is not included in the provided errors.
The error of the spectral indices follows from the Gaussian error propagation
of the flux density errors.

\begin{figure}[bht]
        \resizebox{\hsize}{!}{\includegraphics[trim=1cm 0cm 0.2cm
        0cm]{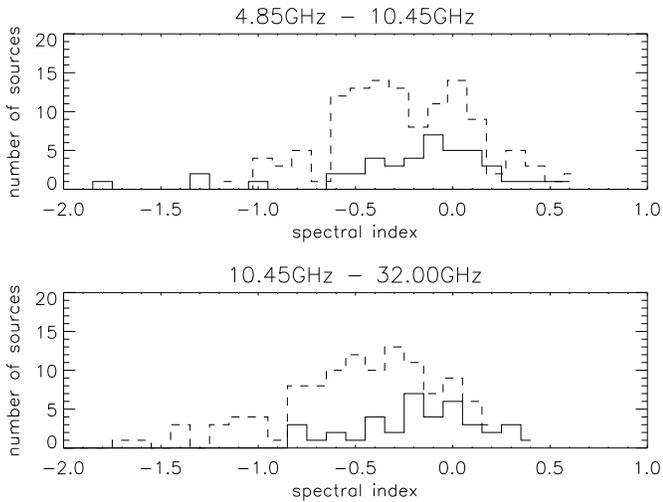}} \caption{The spectral index distribution
        for variable (solid line) and non-variable (dashed line)
        sources. The spectral indices have been calculated from flux
        densities at 4.85~GHz and 10.45~GHz (upper panel) and at
        10.45~GHz and 32.00~GHz (lower panel). }
        \label{fig:specindex}
\end{figure}

The derived distributions of the spectral indices are shown in
Fig.~\ref{fig:specindex}.  The number of sources with a negative
spectral index between 4.85 and 10.45~GHz is 142, that of sources with a
negative spectral index between 10.45~GHz and 32~GHz is 139.  The number
of sources with a positive spectral index between 4.85 and 10.45~GHz is
67 (31\,\%), whereas 41 sources show a positive spectral index between 10.45~GHz
and 32~GHz. Thus, $\sim 20$\% of the sources of our sample show an
inverted spectrum up to 32~GHz. Out of 214 sources 63 (29\,\%) are resolved by
the Effelsberg beam at one or more observing frequencies.
Out of the 63 resolved objects 30 are within $10^\circ$ from the Galactic plane.
Identification statistics about the object types are given in Sect.~\ref{sec:discussion}.

\section{Extended sources and variability \label{sec:ext_var}}

The source spectra extracted with SPECFIND from the database result
from non-simultaneous multi-epoch flux density measurements with
different telescopes. The observations described here, however, have
been obtained with the 100-m Effelsberg telescope, providing for each
source quasi-simultaneous measurements within 15-20~min at 3 different
frequencies.  One expects that radio sources with flat or inverted
spectra show higher degree of variability than those with steeper
spectra. In order to test this idea, we compared the Effelsberg
4.85~GHz (6~cm) flux density to a corresponding 4.85~GHz flux density
in SPECFIND whenever that is available (121 sources of our sample).
The SPECFIND used 4.85~GHz flux densities obtained with angular
resolutions of $3.5'$and $2.8'$ (Gregory et al. 1996; Gregory \& Condon 1991;
Becker et al. 1991; Wright et al. 1994, 1996; Griffith et al. 1994, 1995;
Bennett et al. 1986; Langston et al. 1990, Griffith et al. 1990, 1991), compared to
the Effelsberg 100-m telescope resolution of $2.4'$. Therefore,
extended or confusing sources observed with these different beam sizes
can also lead to differences between the Effelsberg and
the SPECFIND fluxes. 

Since the SPECFIND catalog may contain more than one flux density at
4.85~GHz, we base our variability/confusion criterion on the maximum of the
absolute value of the differences between the SPECFIND and the
Effelsberg 4.85~GHz flux densities $\Delta
S_\mathrm{4.85~GHz}=S_\mathrm{4.85~GHz}^\mathrm{Eff}-S_\mathrm{4.85~GHz}^\mathrm{SPECFIND}$:
\begin{equation}
\max(|\Delta S_\mathrm{4.85~GHz}|)/S^\mathrm{Eff}_\mathrm{4.85~GHz} \geq 0.5\ ,
\label{eq:variable}
\end{equation} 
where $S^\mathrm{Eff}_\mathrm{4.85~GHz}$ is the Effelsberg and
$S^\mathrm{SPECFIND}_\mathrm{4.85~GHz}$ is the SPECFIND flux density
at 4.85~GHz. In addition, we require no overlap between the SPECFIND
and Effelsberg 4.85~GHz points within the errors.
After applying this criterion we find that 59 out of 197
sources are suspected to be variable or confusing. This represents 30\% of the
restricted sample (197 sources having a SPECFIND frequency point at
4.85~GHz) and 28\% of the whole sample (214 sources). Based on these
numbers we expect about 30\% of the sources in our sample to be
variable at centimeter wavelengths. Out of the identified 59 variable
sources 10 are quasars, 4 are BL~Lacertae object, 3 are galaxies, 3 are X-ray
sources, 4 are infrared sources, and 1 is a planetary nebula (PN). As expected, the variable
sources tend to be either BL~Lacs or quasars (see also Torniainen et
al.~2005). 

The information about the source extension and variability can be
found in column~7 of Table~\ref{tab:table1}.
There, the apparently variable/confusing sources based on the 4.85~GHz flux densities
are marked with the symbol 'V'.
The letter `E' indicates sources whose 
extensions are larger than the beam of the Effelsberg telescope at at least one frequency.
We find 30 variable sources which are extended or confusing in at least 
one of our Effelsberg observations (designated as 'EV').
Fig.~\ref{fig:specindex} shows the
spectral index distribution for variable and non-variable
sources. Variable sources have in general a flatter spectral index
between $4.85-10.45$~GHz and $10.45-32$~GHz. We did not test for
variability at 10.45~GHz and 32~GHz, because there are not many
measurements available in the database at these frequencies to compare
with.

%

\section{Cross-identification and GPS spectrum fitting \label{sec:fitting}}

\begin{figure*}[p]
        \center
        \resizebox{14cm}{!}{\includegraphics{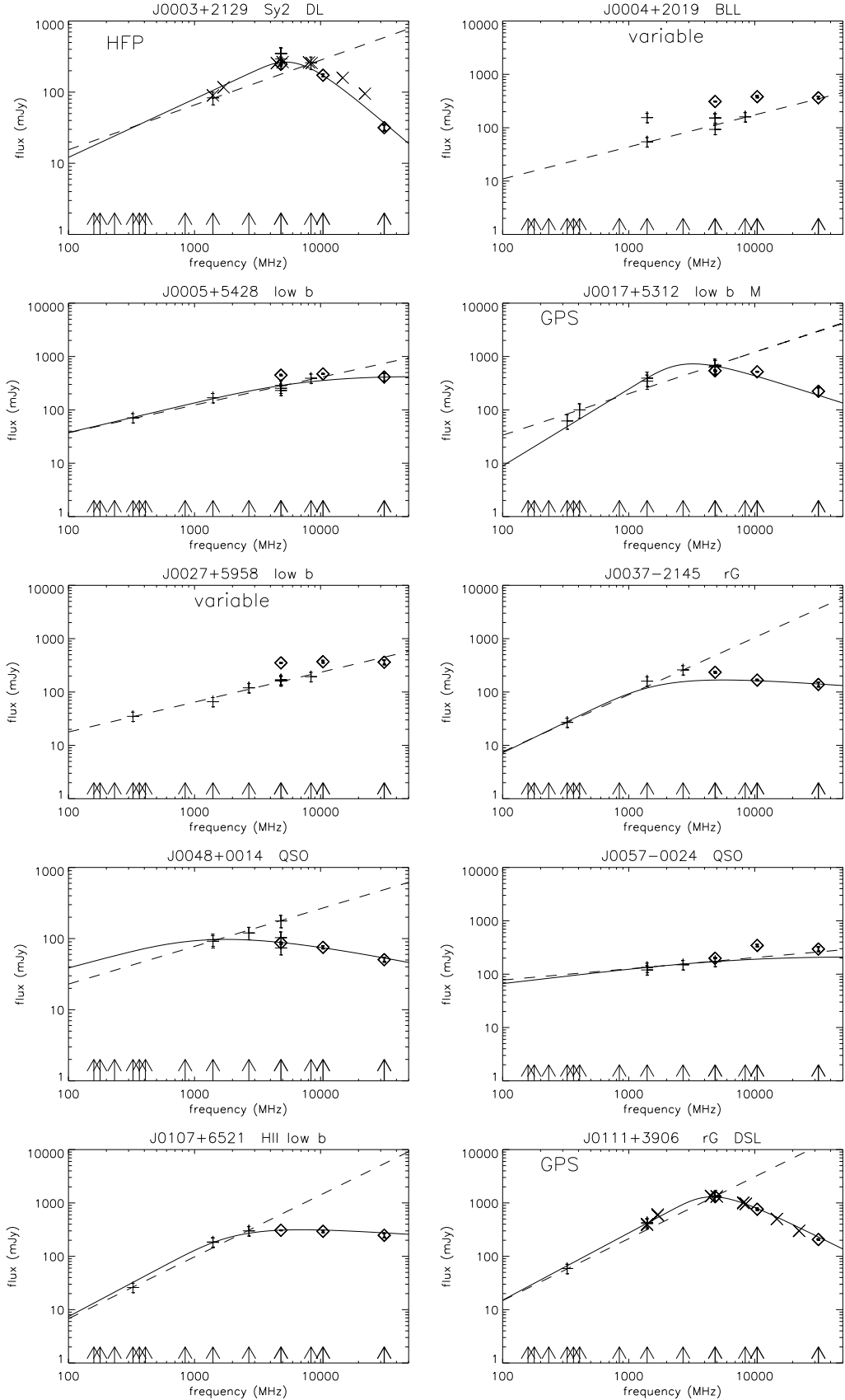}}
        \caption{Radio spectra. SPECFIND data are plotted as pluses with associated
	  error bars, the Effelsberg flux densities as diamonds, and the
	  flux densities obtained by Dallacasa et al. (2000) as crosses.
	  Dashed line: power law fitted by SPECFIND;
	  solid line: fitted spectrum (Eq.~\ref{eq:gpsspec}).
	  GPS: GHz-Peaked Spectrum; HFP: High Frequency Peaker.
	  On top of each panel the source name, object type (if available),
	  and membership of an existing sample (Table~\ref{tab:gpscross}) are given.
	  A source located $\pm 10^{\circ}$ from the Galactic plane is labeled
	  with ``low b''. The arrows on the frequency axis indicate the frequencies
	  of the radio catalogs included in SPECFIND (Vollmer et al. 2005a).
	  {\it The full figure is available in the online version of the Journal.}}
\label{fig:spectra}
\end{figure*}

Fig.~\ref{fig:spectra} shows the combined radio spectra from our
Effelsberg observations (diamonds) and the SPECFIND catalog (plus
signs with error bars) along with the SPECFIND power law spectrum
(dashed line). We cross-identified our sources with the CDS SIMBAD
database (Wenger et al. 2000). The obtained object types are indicated
at the top of each panel and in Table~\ref{tab:table1}.  We have
cross-identified 108 objects: 45 quasars, 7 Seyfert or LINER galaxies, 8
BL~Lacs, 15 galaxies, 8 X-ray sources, 8 infrared
sources, 8 planetary nebulae, 2 stars, and 2 H{\sc ii}
regions. Additionally, we added 5 quasar identifications from V\'eron \&
V\'eron (2006). As one can see in this list, there are 12 confirmed
Galactic sources. As expected, young planetary nebulae and compact
H{\sc ii} regions can also show an inverted radio spectrum up to the
GHz regime. In addition, the infrared sources located at $\pm
10^{\circ}$ from the Galactic plane, which are all extended, are most
probably also of Galactic nature. We end up with 80 confirmed
extragalactic sources with 60\% of them being quasars (from SIMBAD and
V\'eron \& V\'eron~2006).

\begin{figure}[t!]
        \center
        \resizebox{0.9\hsize}{!}{\includegraphics{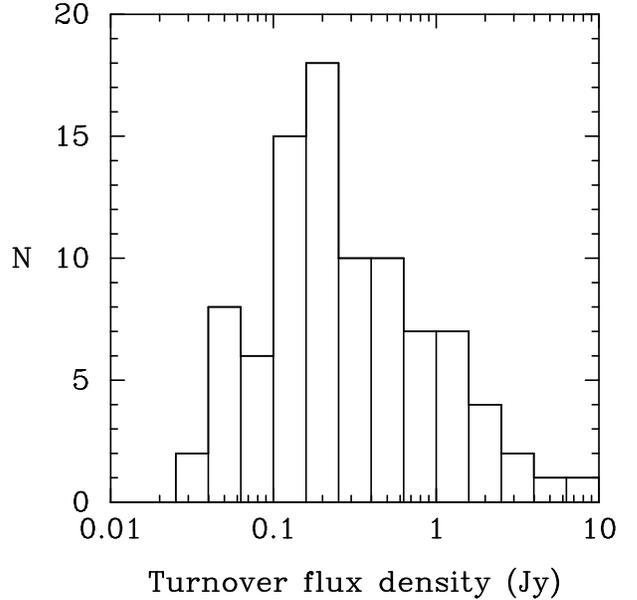}}
        \caption{Distribution of peak flux densities of GPS sources from
        Table~\ref{tab:table1}.
        } \label{fig:turnover}
\end{figure}

Following Snellen et al.~(1998) the combined radio spectra were fitted
by the function:
\begin{equation}
S(\nu)=\frac{S_{\rm max}}{1-{\rm e}^{-1}} \times \big(\frac{\nu}{\nu_{\rm max}}\big)^{k} \times
(1-{\rm e}^{-(\frac{\nu}{\nu_{\rm max}})^{l-k}})\ ,
\label{eq:gpsspec}
\end{equation}
where $k$ is the optically thick spectral index, $l$ the optically
thin spectral index, and $S_{\rm max}$ and $\nu_{\rm max}$
respectively the peak flux density and peak frequency. The optically
thick spectral index $k$ was set to the SPECFIND spectral index 
(slope of the dashed line in Fig.~\ref{fig:spectra}) or to
the spectral index between the two points with the lowest frequencies,
if this allowed a better fit. For the optically thin spectral index,
we used the spectral index between the two points at the highest
frequencies available. The peak flux density and peak frequency were
then determined with a least square fit method. 
We also determine the
FWHM and $\Delta={\rm log}(FWHM\ ({\rm MHz}))$ of the fitted function (Eq.~\ref{eq:gpsspec}). 
A radio source is defined as a GPS if $0 < \Delta \leq 2.5$ and $\nu_{\rm max} \leq 5$~GHz.
Following Dallacasa et al.~(2000) sources with $0 < \Delta \leq 2.5$ and $\nu_{\rm max} > 5$~GHz
or sources with $\alpha > 0.5$ between 1.4 and 5~GHz and $\alpha > 0$ between 
10 and 32~GHz are classified as High Frequency Peakers (HFP).
With these definitions we find 38 GPS and 53 HFP sources. The fitted peak or
turnover frequencies can be found in column~8 of
Table~\ref{tab:table1}. The distribution of peak flux densities is
shown in Fig.~\ref{fig:turnover}.
Most of our sources have a peak flux density around 0.2~Jy. Our GPS/HFP
candidate source sample (Table~\ref{tab:table1}) comprises 18
extended/confusing sources and 4 variable sources (Eq.~\ref{eq:variable}). These
objects are likely blazars rather than genuine young GPS/HFP sources.

\section{Milliarcsecond-scale properties \label{sec:vlbi}}

To investigate parsec-scale characteristics of the sample, we have
measured the VLBI compactness at 2.3 and 8.6~GHz as described in
Kovalev et al. (2005) using the VLBA\footnote{Very
Long Baseline Array of the National Radio Astronomy Observatory}
Calibrator Survey (VCS, Beasley
et al.\ 2002, Fomalont et al.\ 2003, Petrov et al.~2005, 2006, 2008,
Kovalev et al.\ 2007). The compactness is defined as the
ratio of the correlated flux density measured at long interferometer
spacings ($uv$-spacings) and the integrated flux density from a VLBI
image.
Typical errors of the compactness determination are about 0.05.
The error can rise up to 0.1 for the sources with integrated
flux density about or less than 0.1~Jy.

\begin{figure}[t!]
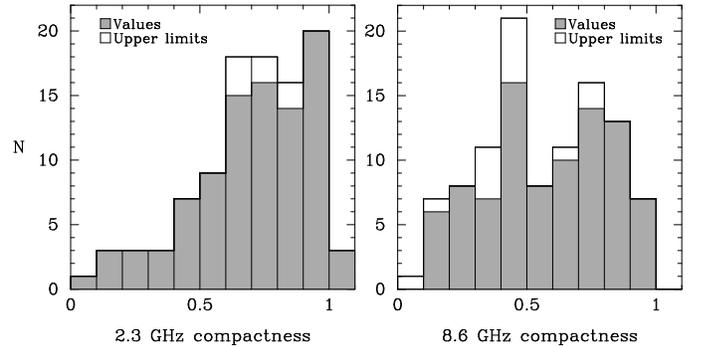

\resizebox{\hsize}{!}{
\includegraphics{hist_Scomp.eps}
\includegraphics[trim=0.6cm 0cm 0cm 0cm,clip]{hist_Xcomp.eps}}
\caption{VLBI compactness  ratios  at  2.3~GHz  (left  panel)  and
8.6~GHz (right panel).}
\label{fig:comp}
\end{figure}

Our sample contains 108 sources from the VLBA Calibrator Survey
(Table~\ref{tab:table1} and Fig.~\ref{fig:comp}).
For further analysis it is convenient to discriminate between almost
point-like and partially resolved sources. We define a source as
(almost) point-like if its compactness is larger than 0.7.
Whereas more than half of the sources are compact at 2.3~GHz, the
distribution of the compactness at 8.6~GHz is approximately flat
(Fig.~\ref{fig:comp}).

%
The majority of HFP candidate sources (21 out of 35) appear point-like at 2.3~GHz
whereas 14 out of 21 GPS candidate sources are extended at this frequency.
Genuine GPS/HFP candidate sources are expected to be resolved at 
both frequencies with compactness values decreasing with increasing 
frequency. We find that 24 out of 53 GPS/HFP candidate sources 
with available compactness information do show this 
behavior. Indeed the median compactness values are 
0.55 (GPS) and 0.75 (HFP) at 2.3 GHz, while at 8.6 GHz they 
decrease to 0.37 (GPS) and 0.44 (HFP).

In a second step, we classified the sources based on their morphology in
the two-frequency VLBI images. This classification is summarized in
column 13 of Table~\ref{tab:table1}: `nc'---naked core candidate;
`cj'---core jet candidate; `cso'---compact symmetric object candidate. A
source is classified here as a cso if distinct components of its
dual-frequency VLBA structure have 2.3--8.6~GHz spectral index
values which differ by less than 0.5.
It is important to note that, on the basis of the two-frequency VLBI
data alone, one cannot definitely decide whether an object is a
classical compact symmetric one. Rather a multi-band VLBI study is
needed in order to reach a definite conclusion (e.g., Peck \& Taylor
2000, Snellen et al.~2000b, Xiang et al.~2002, 2005, 2006, Orienti et
al.~2006). A ``deeper'' VLBI experiment could also detect a weak jet in
many sources classified here as naked cores. In summary, we
identify 65 previously unknown GPS/HFP candidate sources out of which 32
are listed also in the VLBA Calibrator survey catalog.

\begin{figure}
        \resizebox{\hsize}{!}{\includegraphics{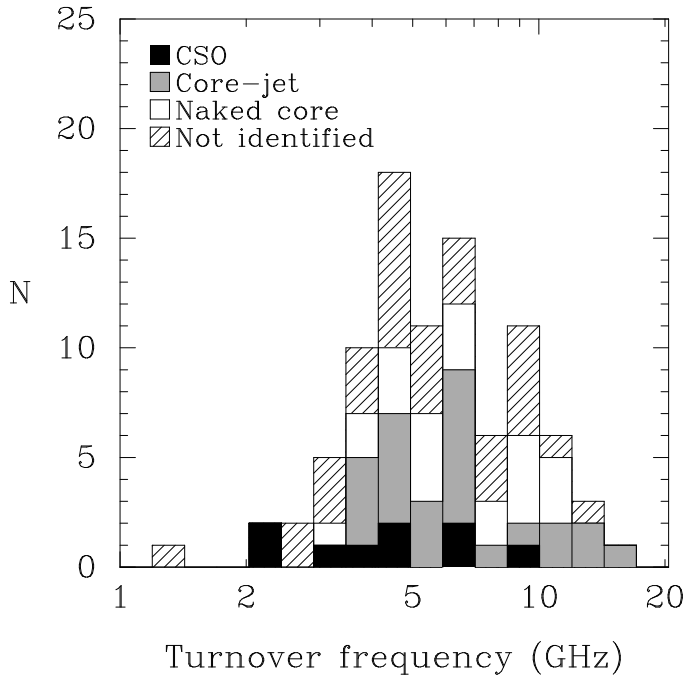}}
        \caption{The distribution of the turnover frequencies for the
        different VLBI morphology classes.}
\label{fig:morph}
\end{figure}

In Fig.~\ref{fig:morph} we show the distribution of the turnover
frequencies for the different VLBI morphology classes. Most of the
sources are point-like (naked cores), about half of this number have
additional jets detected and 9 sources are classified as compact
symmetric object candidates. The CSO candidates which are extended at the
milliarcsecond-scale, tend to have lower values of the turnover frequency in
comparison to the properties of the full sample.

\section{Comparison with other GPS/HFP samples \label{sec:discussion}}

In order to investigate if our objects are already identified as
GPS/HFP sources, we cross-identified our sample with existing GPS/HFP
samples of Snellen et al. (1998), Stanghellini et al. (1998), Marecki
et al. (1999), Dallacasa et al. (2000), Fanti et al. (2001) and
Labiano et al. (2007).  Table~\ref{tab:gpscross} shows the results of
this cross-identification.  The sample designations can also be found
in column~6 of Table~\ref{tab:table1} and on top of each plot in
Fig.~\ref{fig:spectra}.  We find 32 previously known GPS/HFP sources in our
sample, out of which 26 are classified as GPS/HFP candidates by us.
Dallacasa et al. (2000) made quasi-simultaneous multi-frequency
observations of HFP sources at the VLA.  We have plotted their flux
densities as crosses on the spectra of Fig.~\ref{fig:spectra}.  
For the majority of the sources there is a good agreement with
the Effelsberg and SPECFIND flux densities.

\begin{table}
      \caption{Cross-identification with existing GPS/HFP samples.}
         \label{tab:gpscross}
      \[
         \begin{array}{lcc}
          \hline\hline
	 {\rm Article} & {\rm Designation}  & {\rm Number\ of\ objects} \\
	  \hline
	  {\rm Snellen\ et\ al.~(1998)} & {\rm B} & 3 \\
	  {\rm Stanghellini\ et\ al.~(1998)} & {\rm S} & 7 \\
	  {\rm Fanti\ et\ al.~(2001)} & {\rm F} & 0 \\
	 {\rm Dallacasa\ et\ al.~(2000)} & {\rm D} & 15 \\
	 {\rm Marecki\ et\ al.~(1999)} & {\rm M} & 9 \\
	 {\rm Labiano\ et\ al.~(2007)} & {\rm L} & 14 \\
	  \hline
   \end{array}
      \]
\end{table}

\section{Summary \label{sec:summary}}

A sample of 214 radio sources with inverted spectra and declination
$\delta > -25^\circ$ was extracted from the SPECFIND catalog (Vollmer et
al. 2005a). This catalog contains cross-identifications of radio sources
from surveys at different frequencies and combines them into one single
radio spectrum per object. To obtain quasi-simultaneous radio spectra,
we observed those sources with the MPIfR Effelsberg 100-m radio
telescope at 4.85, 10.45, and 32~GHz. All the observed sources were
detected at 4.85~GHz, while 209 and 180 sources were detected at
10.45~GHz and 32~GHz, respectively. We expect the fraction of variable
or confusing sources in our complete sample to be of $\sim 30$\%. On the basis of
the performed analysis of continuum radio spectra, we have identified 38 GPS
and 53 HFP candidates out of which 65 were previously unknown.
An inspection of VCS data shows that 24 out of 53 GPS/HFP sources
with available VCS data are resolved at 2.3 and 8.6~GHz. We have confirmed the expected 
tendency for HFP objects to be highly compact and the GPS ones to show a significantly
lower level of compactness at the milliarcsecond scale. This independently
supports robustness of our source classifications presented here.

%
%

We have a success of $\sim$45\% for finding GPS/HFP candidates from a
selected SPECFIND catalog sample. Once the SPECFIND catalog is upgraded
with the inclusion of more radio catalogs, this method comprises a
promising way for future identification of new GPS/HFP sources.

\begin{acknowledgements}

This work is based on observations with the 100-m telescope of the MPIfR
(Max-Planck-Institut f\"{u}r  Radioastronomie) at Effelsberg. We would
like to thank A.~Kraus for his help. Y.~Y.~Kovalev is a Research Fellow
of the Alexander von Humboldt Foundation. We would like to thank John
McKean for fruitful discussions.
We thank the anonymous referee for thoughtful reading and
useful comments which helped to improve the manuscript.

\end{acknowledgements}

\begin{table*}
      \caption{Results of the observations {\it The full table is available in electronic form at the CDS.} \label{tab:table}}
      \begin{center}
      \[
\begin{array}{lccccrrc}
           \hline\hline
           \noalign{\smallskip}
           {\rm Name\ (J2000)} & {\rm RA,\ DEC\ (J2000)} & \multicolumn{3}{c}{\rm{}Flux~density~(mJy)} & \multicolumn{2}{c}{\rm{}Spectral~index} & {\rm Resolved} \\
            & {\rm (HHMMSS.ss)\ (DDMMSS.s)} & {\rm 4.85~GHz} & {\rm 10.45~GHz}  & {\rm 32.0~GHz} & 4.85/10.45~{\rm GHz} & 10.45/32.00~{\rm GHz} &  ({\rm GHz}) \\
           \hline
  {\rm J}0003+2129 & 	00 03 19.35	~+21 29 44.4 &        247\pm         2 &        173\pm         7 &         32\pm         3 & -0.45\pm 0.05 & -1.51\pm 0.09 & \\
  {\rm J}0004+2019 & 	00 04 35.76	~+20 19 42.2 &        310\pm         2 &        382\pm        14 &        365\pm        19 &  0.27\pm 0.05 & -0.03\pm 0.06 & \\
  {\rm J}0005+5428 & 	00 05 04.37	~+54 28 24.9 &        448\pm         9 &        472\pm         3 &        413\pm        71 &  0.07\pm 0.03 & -0.11\pm 0.15 & \\
  {\rm J}0017+5312 & 	00 17 51.70	~+53 12 19.0 &        541\pm        11 &        518\pm         4 &        223\pm        38 & -0.05\pm 0.03 & -0.74\pm 0.15 & \\
  {\rm J}0027+5958 & 	00 27 03.29	~+59 58 53.0 &        352\pm         2 &        370\pm        18 &        362\pm        34 &  0.06\pm 0.06 & -0.01\pm 0.09 & \\
  {\rm J}0037-2145 & 	00 37 14.79	~-21 45 24.5 &        235\pm         7 &        167\pm         4 &        139\pm        13 & -0.43\pm 0.05 & -0.16\pm 0.09 & \\
\hline
        \end{array}
      \]
      \end{center}
\end{table*}

\addtocounter{table}{-1}
\begin{table*}
      \caption{Results of the observations (continued)}
      \begin{center}
      \[
         \begin{array}{lccccrrc}
           \hline\hline
           \noalign{\smallskip}
           {\rm Name\ (J2000)} & {\rm RA,\ DEC\ (J2000)} & \multicolumn{3}{c}{\rm{}Flux~density~(mJy)} & \multicolumn{2}{c}{\rm{}Spectral~index} & {\rm Resolved} \\
            & {\rm (HHMMSS.ss)\ (DDMMSS.s)} & {\rm 4.85~GHz} & {\rm 10.45~GHz}  & {\rm 32.0~GHz} & 4.85/10.45~{\rm GHz} & 10.45/32.00~{\rm GHz} &  ({\rm GHz}) \\
           \hline
  {\rm J}0703-0517 & 	07 03 08.44	~-05 17 17.5 &        128\pm         3 &        190\pm         5 &        135\pm         4 &  0.52\pm 0.04 & -0.29\pm 0.03 & \\
  {\rm J}0708+3524 & 	07 08 49.56	~+35 24 22.0 &         60\pm         2 &         52\pm         2 &          - & -0.18\pm 0.05 &  -  & 5 \\
  {\rm J}0713+4349 & 	07 13 38.10	~+43 49 17.0 &       1543\pm        77 &        973\pm       119 &        398\pm        89 & -0.59\pm 0.17 & -0.79\pm 0.23 & \\
  {\rm J}0717-1319 & 	07 17 11.52	~-13 19 04.7 &         36\pm         2 &         26\pm        26 &         22\pm        19 & -0.42\pm 1.31 & -0.15\pm 1.19 & 10 \\
  {\rm J}0717+5231 & 	07 17 29.99	~+52 31 04.9 &        124\pm         2 &         85\pm         2 &         43\pm         5 & -0.49\pm 0.04 & -0.59\pm 0.10 & 5 \\
  {\rm J}0722+3722 & 	07 22 01.20	~+37 22 28.0 &        201\pm        10 &        154\pm        19 &         65\pm        15 & -0.34\pm 0.17 & -0.76\pm 0.23 & \\
\hline
        \end{array}
      \]
      \end{center}
\end{table*}

\addtocounter{table}{-1}
\begin{table*}
      \caption{Results of the observations (continued)}
      \begin{center}
      \[
         \begin{array}{lccccrrc}
           \hline\hline
           \noalign{\smallskip}
           {\rm Name\ (J2000)} & {\rm RA,\ DEC\ (J2000)} & \multicolumn{3}{c}{\rm{}Flux~density~(mJy)} & \multicolumn{2}{c}{\rm{}Spectral~index} & {\rm Resolved} \\
            & {\rm (HHMMSS.ss)\ (DDMMSS.s)} & {\rm 4.85~GHz} & {\rm 10.45~GHz}  & {\rm 32.0~GHz} & 4.85/10.45~{\rm GHz} & 10.45/32.00~{\rm GHz} &  ({\rm GHz}) \\
           \hline
  {\rm J}1359+5544 & 	13 59 05.75	~+55 44 29.3 &         84\pm         2 &        100\pm         2 &        142\pm        12 &  0.22\pm 0.04 &  0.31\pm 0.08 & \\
  {\rm J}1359+4011 & 	13 59 38.09	~+40 11 38.3 &        314\pm         6 &        347\pm        42 &        258\pm       106 &  0.13\pm 0.16 & -0.25\pm 0.38 & \\
  {\rm J}1407+2827 & 	14 07 00.30	~+28 27 14.0 &       2572\pm        47 &       1469\pm       179 &        234\pm        96 & -0.72\pm 0.16 & -1.63\pm 0.38 & \\
  {\rm J}1408-1529 & 	14 08 36.08	~-15 29 24.4 &         50\pm         2 &         46\pm         2 &         43\pm         8 & -0.10\pm 0.09 & -0.03\pm 0.17 & 5 \\
  {\rm J}1410+0731 & 	14 10 35.08	~+07 31 21.5 &        185\pm         4 &        198\pm        24 &        174\pm        72 &  0.09\pm 0.16 & -0.10\pm 0.38 & \\
  {\rm J}1420+1703 & 	14 20 20.89	~+17 03 29.2 &        276\pm         5 &        445\pm        54 &        423\pm       174 &  0.62\pm 0.16 & -0.04\pm 0.38 & \\
\hline
        \end{array}
      \]
      \end{center}
\end{table*}

\addtocounter{table}{-1}
\begin{table*}
      \caption{Results of the observations (continued)}
      \begin{center}
      \[
         \begin{array}{lccccrrc}
           \hline\hline
           \noalign{\smallskip}
           {\rm Name\ (J2000)} & {\rm RA,\ DEC\ (J2000)} & \multicolumn{3}{c}{\rm{}Flux~density~(mJy)} & \multicolumn{2}{c}{\rm{}Spectral~index} & {\rm Resolved} \\
            & {\rm (HHMMSS.ss)\ (DDMMSS.s)} & {\rm 4.85~GHz} & {\rm 10.45~GHz}  & {\rm 32.0~GHz} & 4.85/10.45~{\rm GHz} & 10.45/32.00~{\rm GHz} &  ({\rm GHz}) \\
           \hline
  {\rm J}2016+6450 & 	20 16 40.66	~+64 50 28.1 &        250\pm         4 &        245\pm        14 &        206\pm         7 & -0.01\pm 0.08 & -0.15\pm 0.06 & \\
  {\rm J}2018-1109 & 	20 18 27.99	~-11 09 55.1 &        363\pm        18 &        216\pm         5 &          - & -0.67\pm 0.07 &  -  & \\
  {\rm J}2022+6136 & 	20 22 06.68	~+61 36 58.8 &       3034\pm        28 &       2714\pm       115 &       1347\pm        23 & -0.14\pm 0.06 & -0.62\pm 0.04 & \\
  {\rm J}2022+3519 & 	20 22 35.65	~+35 19 36.8 &         44\pm         2 &         25\pm         2 &         27\pm         5 & -0.72\pm 0.10 &  0.08\pm 0.18 & 5/10 \\
  {\rm J}2026+3903 & 	20 26 11.61	~+39 03 32.8 &        116\pm         3 &         92\pm         2 &         51\pm         9 & -0.30\pm 0.05 & -0.52\pm 0.16 & 32 \\
  {\rm J}2032-2125 & 	20 32 54.29	~-21 25 16.6 &         91\pm         5 &         23\pm         2 &          - & -1.75\pm 0.11 &  -  & \\
\hline			
        \end{array}			
      \]				
      \end{center}			
\end{table*}				
					
\begin{table*}				
      \caption{Properties of the sources. {\it The full table is available in electronic form at the CDS.}}
      \label{tab:table1}
      \begin{center}
      \[
         \begin{array}{llccccccccccc}
           \hline\hline
           \noalign{\smallskip}
	   {\rm Name\ (J2000)} & {\rm redshift} & {\rm VCS} & {\rm type} & {\rm GPS/HFP} & {\rm sample} & {\rm var./ext.} & \nu_{\rm peak}\ {\rm (MHz)} & {\rm S_{\rm peak}}\ {\rm (mJy)} & {\rm note} & c_{8.6{\rm GHz}} & c_{2.3{\rm GHz}}& {\rm class.} \\
	   \hline
  {\rm J}0003+2129 &  0.450 & {\rm y} & {\rm   Sy2} & {\rm HFP} & {\rm   DL} & {\rm  } &   6800 & 246 & {\rm      } & 0.71 & 0.64 & {\rm cso} \\
  {\rm J}0004+2019 &  0.677 & {\rm y} & {\rm   BLL} & {\rm    } & {\rm     } & {\rm V} &        & & {\rm      } & 0.25 & 0.92 & {\rm nc}\\
  {\rm J}0005+5428 &        & {\rm y} & {\rm      } & {\rm } & {\rm     } & {\rm  } &       & & {\rm  low\,b} & 0.71 & 0.89 & {\rm nc}\\
  {\rm J}0017+5312 &        & {\rm y} & {\rm      } & {\rm GPS} & {\rm    M} & {\rm  } &   2900 & 721 & {\rm  low\,b} & 0.26 & 0.28 & {\rm cj/cso} \\
  {\rm J}0027+5958 &        & {\rm y} & {\rm      } & {\rm    } & {\rm     } & {\rm V} &       & & {\rm  low\,b} &0.86 & <0.72 & {\rm nc}\\
  {\rm J}0037-2145 &        & {\rm y} & {\rm    rG} & {\rm    } & {\rm     } & {\rm  } &       & & {\rm      } &0.45 & 0.75 & {\rm cj}\\
  {\rm J}0048+0014 &  1.536 & & {\rm   QSO} & {\rm    } & {\rm     } & {\rm  } &        & & {\rm      } & & & \\
\hline
        \end{array}
      \]
      \end{center}
\end{table*}

\addtocounter{table}{-1}
\begin{table*}
      \caption{Properties of the sources (continued)}
      \begin{center}
      \[
         \begin{array}{llccccccccccc}
           \hline\hline
           \noalign{\smallskip}
	   {\rm Name\ (J2000)} & {\rm redshift} & {\rm VCS} & {\rm type} & {\rm GPS/HFP} & {\rm sample} & {\rm var./ext.} & \nu_{\rm peak}\ {\rm (MHz)} & {\rm S_{\rm peak}}\ {\rm (mJy)} & {\rm note} & c_{8.6{\rm GHz}} & c_{2.3{\rm GHz}}& {\rm class.}\\
	   \hline
  {\rm J}0703-0517 &        & & {\rm      } & {\rm HFP} & {\rm     } & {\rm  } &   8700 & 142 & {\rm  low\,b} & & & \\
  {\rm J}0708+3524 &        & & {\rm      } & {\rm    } & {\rm     } & {\rm E} &      &  & {\rm      } & & & \\
  {\rm J}0713+4349 &  0.518 & {\rm y} & {\rm   EmG} & {\rm GPS} & {\rm   SL} & {\rm  } &   2100 & 2233 & {\rm      } & 0.14 & 0.35 & {\rm cso}\\
  {\rm J}0717-1319 &        & & {\rm      } & {\rm    } & {\rm     } & {\rm EV} &      &  & {\rm  low\,b} & & & \\
  {\rm J}0717+5231 &        & & {\rm      } & {\rm GPS} & {\rm     } & {\rm E} &   3400 & 130 & {\rm      } & & & \\
  {\rm J}0722+3722 &  1.629 & {\rm y} & {\rm   QSO} & {\rm  GPS} & {\rm   DM} & {\rm V} &  4200 & 223  & {\rm      } & 0.33 & 0.91 & {\rm cj} \\
  {\rm J}0728+5907 &        & & {\rm      } & {\rm HFP} & {\rm     } & {\rm  } &   8700 & 50 & {\rm      } & & & \\
\hline
        \end{array}
      \]
      \end{center}
\end{table*}

\addtocounter{table}{-1}
\begin{table*}
      \caption{Properties of the sources (continued)}
      \begin{center}
      \[
         \begin{array}{llccccccccccc}
           \hline\hline
           \noalign{\smallskip}
	   {\rm Name\ (J2000)} & {\rm redshift} & {\rm VCS} & {\rm type} & {\rm GPS/HFP} & {\rm sample} & {\rm var./ext.} & \nu_{\rm peak}\ {\rm (MHz)} & {\rm S_{\rm peak}}\ {\rm (mJy)} & {\rm note} & c_{8.6{\rm GHz}} & c_{2.3{\rm GHz}}& {\rm class.}\\
	   \hline
  {\rm J}1359+5544 &        & {\rm y} & {\rm      } & {\rm    } & {\rm     } & {\rm V} &     &   & {\rm      } & & 0.57 & {\rm nc} \\
  {\rm J}1359+4011 &  0.407 & {\rm y} & {\rm   EmG} & {\rm    } & {\rm    M} & {\rm  } &     &   & {\rm      } & 0.85 & 1.00 & {\rm nc}  \\
  {\rm J}1407+2827 &  0.077 & {\rm y} & {\rm   Sy1} & {\rm HFP} & {\rm   DSL} & {\rm  } &   5900 & 2550 & {\rm      } & 0.42 & 0.48 & {\rm cso}\\
  {\rm J}1408-1529 &        & & {\rm     X} & {\rm    } & {\rm     } & {\rm EV} &    &    & {\rm      } & & & \\
  {\rm J}1410+0731 &        & {\rm y} & {\rm      } & {\rm    } & {\rm    } & {\rm V} &     &   & {\rm      } & 0.51 & 0.58 &{\rm cj} \\
  {\rm J}1420+1703 &        & {\rm y} & {\rm      } & {\rm } & {\rm     } & {\rm  } &    &    & {\rm      } & 0.89 & & {\rm nc} \\
  {\rm J}1421+4645 &  1.665 & {\rm y} & {\rm   QSO} & {\rm } & {\rm     } & {\rm  } &    &   & {\rm      } & 0.68 & 0.95 & {\rm nc} \\
\hline
        \end{array}
      \]
      \end{center}
\end{table*}

\addtocounter{table}{-1}
\begin{table*}
      \caption{Properties of the sources (continued)}
      \begin{center}
      \[
         \begin{array}{llccccccccccc}
           \hline\hline
           \noalign{\smallskip}
	   {\rm Name\ (J2000)} & {\rm redshift} & {\rm VCS} & {\rm type} & {\rm GPS/HFP} & {\rm sample} & {\rm var./ext.} & \nu_{\rm peak}\ {\rm (MHz)} & {\rm S_{\rm peak}}\ {\rm (mJy)} & {\rm note} & c_{8.6{\rm GHz}} & c_{2.3{\rm GHz}}& {\rm class.}\\
	   \hline
  {\rm J}2016+6450 &        & & {\rm      } & {\rm } & {\rm     } & {\rm  } &     &  & {\rm      } & & & \\
  {\rm J}2018-1109 &        & {\rm y} & {\rm      } & {\rm HFP} & {\rm     } & {\rm  } &   6500 & 249 & {\rm      } & 0.44 & 0.92 & {\rm cj}\\
  {\rm J}2022+6136 &  0.227 & {\rm y} & {\rm   Sy2} & {\rm GPS} & {\rm     } & {\rm  } &   4900 & 3050 & {\rm      } & 0.25 & 0.18 & {\rm cso} \\
  {\rm J}2022+3519 &        & & {\rm      } & {\rm    } & {\rm     } & {\rm EV} &     &   & {\rm  low\,b} & & & \\
  {\rm J}2026+3903 &        & & {\rm    IR} & {\rm HFP} & {\rm     } & {\rm E} &   5500 & 105 & {\rm  low\,b} & & & \\
  {\rm J}2032-2125 &        & {\rm y} & {\rm      } & {\rm    } & {\rm     } & {\rm V} &     &   & {\rm      } & & 0.88 & {\rm nc} \\
  {\rm J}2049+1003 &        & {\rm y} & {\rm      } & {\rm HFP} & {\rm     } & {\rm  } &   6300 & 527     & {\rm      } & 0.44 & 0.54 & {\rm cj}\\
\hline
        \end{array}
      \]
      \end{center}

Col.~(1): source name; col.~(2): redshift from V\'eron-Cetty \& V\'eron (2006);
col.~(3): VCS data `y'= available; col.~(4): object type from SIMBAD
(rG: radio galaxy, IR: infrared source, GiC: galaxy in cluster; *iC: star in cluster, 
EmG: emisson line galaxy, PaG: Pair of galaxies, QSO: quasar, Sy: Seyfert galaxy,
BLL: BL Lac-type object, X: X-ray source, PN: planetary nebula, *: star, G: galaxy,
HII: HII region); col.~(5): GPS/HFP,
the classification criteria are described in Sect.~\ref{sec:fitting};
col.~(6): sample membership (abbreviation as in Tab.~\ref{tab:gpscross});
col.~(7): `V'=variable, `E'=extended at 4.85~GHz; col.~(8): peak
frequency in MHz; col.~(9): peak flux density in mJy; 
col.~(10): low\,b=$\pm 10^{\circ}$ from the Galactic plane, `ext?'=
source size $> 45''$;
col.~(11): compactness ratio at 8.6~GHz;
col.~(12): compactness ratio at 2.3~GHz;
col.~(13): classification based on VLBI images: `nc'= naked core candidate;
`cj'= core-jet candidate; `cso'= compact symmetric object candidate;
'cj/cso' = core-jet or compact symmetric object candidate,
components cross-identification between 2 and 8~GHz is questionable;
the VCS classification for the following sources was corrected according
to deeper multi-frequency VLBI studies: J0003+2129 (Orienti et al.~2006),
1407+2827 (e.g., Stanghellini et al.~1997), J1551+5806 (Snellen et al.~2000);
our cso candidates J0111+3906, J0713+4349, J1823+7938, J2022+6136 are confirmed by Peck \& Taylor (2000).

\end{table*}
  
\end{document}